# Sunspot Numbers and Areas from the Madrid Astronomical Observatory (1876-1986)

A.J.P. Aparicio[1] • J.M. Vaquero[1] • V.M.S. Carrasco[1,2] • M.C. Gallego[1]

[1] Departamento de Física, Universidad de Extremadura, Spain

[2] Centro de Geofísica de Évora, Universidade de Évora, Portugal

**Abstract**

The solar program of the Astronomical Observatory of Madrid started in 1876. For ten solar cycles, observations were made in this institution to determine sunspot numbers and areas. The program was completed in 1986. The resulting data have been published in various Spanish scientific publications. The metadata allowed four periods of this program (with different observers and instruments) to be identified. In the present work, the published data were retrieved and digitized. Their subsequent analysis showed that most of these data can be considered reliable given their very high correlation with international reference indices (International Sunspot Number, Group Sunspot Number, and Sunspot Area). An abrupt change emerged in the spots/groups ratio in 1946 which lasted until 1972.

## 1. Introduction

The sunspot number (SN) forms the primary time series in solar and solar-terrestrial physics. Astrophysicists, solar physicists, and geoscientists interested in long-term change in the Sun use it extensively (Hoyt & Schatten, 1998; Vaquero, 2007; Clette, 2011). However, recent work has shown there to be some problems with the quality and homogeneity of this index (Svalgaard, 2012; Vaquero, Trigo & Gallego, 2012; Cliver, Clette & Svalgaard, 2013). Thus, there is real interest in retrieving long series of sunspot records to contribute to improving knowledge of the quality and homogeneity of SN data (e.g., Carrasco et al., 2013).

The Astronomical Observatory of Madrid (AOM) was founded in the late 18th century. Systematic solar observations began in 1876 and continued on a regular basis, although with some gaps, until 1986. Thus, the SN was recorded in this institution for ten consecutive solar cycles. In addition, the astronomers at Madrid measured and recorded the sunspot area in various periods of the 20th century, including from 1952 to 1986 without interruption.

The purpose of the present work is to recover the AOM sunspot number and sunspot area series, to verify their quality, and to compare them with similar, well-known series. The paper is structured as follows: Section 2 describes the metadata that are available for these solar observations. The AOM SN data are examined in Section 3. An abrupt change of the spots/groups ratio emerged from the data and is considered in detail in Section 4. Sunspot area measurements are discussed in Section 5. Finally, Section 6 presents the conclusions of the study.

**2. Metadata**

Thanks to the abundance of published information (see Appendix A; López Arroyo, 2004; Ruiz-Castell, 2008), one has available interesting metadata about the AOM solar program's instruments, methods, and observers. In the following subsections, we shall summarize the information in this metadata.

*First period (1868-1896): V. Ventosa*

The first sunspot observations in the AOM were made in 1868. The observer was Vicente Ventosa, who started to work on solar physics in that year. Ventosa only began to publish his sunspot observations systematically in 1876. The observations of the first years are lost. Ventosa made these observations with a Merz equatorial refractor telescope that had been installed in the AOM in 1858 (D = 27 cm aperture and f = 4.87 m focal length). The image was projected onto a screen, with a solar disk diameter slightly less than 20 cm. To record the observations, Ventosa used a cardboard sheet on which was printed a circle the size of the solar image. The solar program was interrupted in 1897 before Ventosa's retirement.

*Second period (1906-1923): M. Aguilar*

Six years later, in 1903, Fernández Soler & Reig Ascarza (1903) made some non-systematic solar observations (López Arroyo, 2004). In 1906, Miguel Aguilar Cuadrado re-initiated systematic sunspot observations at the Observatory. He could no longer use the Merz telescope because the cylindrical dome that protected it had been dismantled in the meantime (due to mechanical problems), and a new one had yet to be installed. Instead he observed visually with a Steinheil equatorial telescope (D = 12 cm, f = 1.85 m) which had been acquired in 1860. This telescope was installed in the south-west area of the observatory, in a small cylindrical building covered by a modest iron and canvas dome.

In 1909, Aguilar began to make photographic observations. To this end, he used the same Steinheil equatorial telescope and a Zeiss camera (equipped with a magnifying lens that enlarged the 2 cm focal plane image to one 15 cm in diameter). With a sector shutter, exposure times could be as short as 1/500 s. Thus, to obtain correctly exposed images, he just needed to stop down the lens slightly and interpose a violet filter.

Because of lack of experience, unawareness of the proper use of the instruments, and various other difficulties, the resulting collection of negatives lacked the uniformity needed for sunspot area studies. Subsequently, these difficulties were overcome, but by then a prolonged period of years of low solar activity had begun. As a consequence, observers did not consider it appropriate to publish sunspot area records. From 1914 onwards, interest in solar activity increased, and maintaining sunspot area records began in earnest.

But due to various health problems and failure of the camera's shutter, Aguilar began to observe less regularly. Indeed, for the time span 1921-1923, he published only short annual summaries. With his death in 1925, the AOM solar program was once again interrupted.

### *Third period (1931-1933): R. Carrasco*

In November 1931, Rafael Carrasco Garrorena resumed the sunspot observations. At first, he was assisted by Miguel Aguilar Stuyck. In mid-1932, Aguilar Stuyck was replaced by Marcelo Santaló Sors, who left the Observatory in May 1933. Thereafter, Carrasco carried out the task alone. He used the same equipment as Aguilar – the Steinheil equatorial telescope with the Zeiss photoheliograph.

### *Fourth period (1935-1986): E. Gullón and others*

Enrique Gullón de Senespleda began to observe the Sun in 1935. He would be responsible for this task for more than 30 years. He took no solar photographs, but made visual observations by projecting the solar image (drawing the sunspots). He used a Grubb equatorial telescope which the Observatory had purchased for the observation of the total solar eclipse of 1900 (D = 20 cm, f = 3 m), and a Herschel helioscope eyepiece to project a 25 cm diameter image onto a screen (López Arroyo, 2004).

Because of the Spanish Civil War, the observations for the years 1937 and 1938 were made in Valencia (on Spain's east coast). A Zeiss telescope (D = 15 cm, f = 2.20 m) with a Herschel helioscope eyepiece, belonging to the University of Valencia, was used in these observations. Observations were suspended during most of 1939, but in September 1939, Gullón resumed the observations at the Madrid Observatory with the collaboration of Martin Loron. For these observations, the Grubb telescope was again used (López Arroyo, 2004).

Gullón died in 1969. José Iglesias then became the main observer. He followed the same observing methodological approach as his predecessor. Later, in 1973, Manuel López Arroyo became responsible for this task, and Angel Gil was the last observer, from 1984 to 1986 (López Arroyo, 2004).

Figure 1 shows the temporal distribution of the observers during this last period. They were many in number, and there were many observing days per year (which, for some years, reached exceptional values of more than 300 days).

*Recovered data*

The data used for this study were retrieved from the data published by R. Wolf and from two journals – the *Anuario del Observatorio de Madrid* and the *Boletín Astronómico del Observatorio de Madrid*. The Appendix gives a detailed list of all the original sources consulted. We retrieved and kept all the data available giving the sunspot number, the group number, and the sunspot area. Table 1 gives the distribution over time of the data retrieved. As can be seen, the data was grouped into four periods, each of which has its own type of available data and frequency of reporting. To access these original sources, we consulted the archives and libraries of the following institutions: National Astronomical Observatory (Madrid, Spain), Astronomical Observatory of Lisbon (Lisbon, Portugal), Royal Observatory of the Spanish Navy (San Fernando, Spain), and Territorial Centre AEMET (Badajoz, Spain). We then digitized these data, with the result now being available as supplementary material of this paper and in the "Historical Archive of Sunspot Observations" (http://haso.unex.es).

**3. Sunspot Number**

In the present study, monthly means are calculated from daily data, and yearly means from monthly data (the monthly means from daily records, or else monthly records, depending on availability). By way of example, Table 2 gives the yearly group count means for the entire study period and the number of days observed for each year.

From the group number $g$ and the spot number $s$, we define the Madrid Sunspot Number, $MSN = 10\,g + s$, and the Madrid Group Sunspot Number, $MGSN = 12.08\,g$, according to the classical definitions (Cliver, Clette and Svalgaard, 2013).

Figure 2 shows the temporal evolution of these indices. The *MGSN* series is closer to complete than the *MSN* series because group counts were recorded throughout the study period, while spot counts were made in fewer years (see Table 1). Note that, although *MSN* is greater than *MGSN*, the shapes of the cycles are roughly the same in the two series, reaching their maximum and minimum values in the same years except for cycles 5 and 7, in which *MGSN* peaked one year after *MSN*.

In order to analyse the reliability of the Madrid indices, we studied their relationships with international indices. For this purpose, we made linear least-square fits, taking as dependent variables the monthly Madrid indices and as independent variables the corresponding monthly international indices. These latter indices were the International Sunspot Number (*ISN*) for *MSN*, and the Group Sunspot Number (*GSN*) for *MGSN*. For *MSN* (see the left panel of Figure 3), we found a very strong correlation for the first period ($r = 0.972$), which decreased for the last period ($r = 0.918$). For the whole period, the correlation was still strong ($r = 0.925$). For *MGSN* (see the right panel of Figure 3), we obtained very high correlation coefficients for the periods 1876-1896 and 1935-1986 ($r = 0.962$ and $0.961$, respectively), but lower for 1906-1920 and 1931-1933 ($r = 0.908$ and $0.847$, respectively). For the whole period, we also found a strong correlation ($r = 0.931$). The statistical significance corresponded to a confidence level greater than 99.9% for all the aforementioned fits. One can thus take *MSN* and *MGSN* to be reliable indices given their very strong correlations with their respective international indices.

The following are the equations of the linear fits corresponding to the different periods (and the total period):

| | |
|---|---|
| 1876-1896: | $MSN = (1.54 \pm 0.02)ISN + (2.6 \pm 1.1)$ |
| 1936-1986: | $MSN = (1.33 \pm 0.02)ISN + (-1.9 \pm 2.2)$ |
| 1876-1986: | $MSN = (1.31 \pm 0.02)ISN + (2.7 \pm 1.5)$ |
| 1876-1896: | $MGSN = (0.99 \pm 0.02)GSN + (6.5 \pm 0.8)$ |
| 1906-1920: | $MGSN = (0.95 \pm 0.03)GSN + (3.8 \pm 1.9)$ |
| 1931-1933: | $MGSN = (1.39 \pm 0.18)GSN + (-1.1 \pm 2.3)$ |
| 1935-1986: | $MGSN = (0.76 \pm 0.01)GSN + (0.3 \pm 0.8)$ |
| 1876-1986: | $MGSN = (0.75 \pm 0.01)GSN + (6.4 \pm 0.7)$ |

**4. An abrupt change in the spots/groups ratio at around 1946**

In this section, we shall examine the temporal evolution of the spots/groups ratio. For this, we use the indices studied above since their ratio $MSN/MGSN = (1/1.208) + (s/12.08g)$ is itself an indicator of the spots/groups ratio. We calculated monthly values of *MSN*/*MGSN* from the ratio of the monthly means of *MSN* and *MGSN*, and then calculated yearly values from the ratio of the yearly means of *MSN* and *MGSN*. Note that this analysis was only applicable to the first and fourth periods because, although all the periods have group records, only the first and fourth have spot records (see Table 1).

The international ratio varies around unity (Figure 4), but the yearly Madrid ratio is always greater than unity, as is the monthly ratio almost always. This was to be expected since the Madrid indices are not multiplied by their correction factors. However, if, as statistical studies indicate on average (Waldmeier, 1968), one applies the condition $s = 10g$ to the above expression for *MSN/MGSN*, one obtains a value of 1.66, a value around which the Madrid ratio varies. Therefore, although the two series are not similar in value, they each vary around their expected value. This can be checked in another way. Note that $1/1.66 = 0.6$, which is the scaling factor applied to the modern *ISN* to match the original Wolf values. Therefore, the Madrid series (on average) closely matches the international reference series (without the 0.6 correction factor).

Examining the Madrid series more closely, one sees that their ratio is more or less stable in value until 1946, when it undergoes an abrupt rise followed by major oscillations which extend until 1972. From 1973 onwards, the ratio returns to values close to those before 1946. This "unstable phase" (1946-1972) is the one with the greatest standard deviations. This is because, in this phase, the yearly means are constructed using monthly data with greater scatter. This average rise in the ratio for the time span 1946-1972 occurs not only in the Madrid series. The international ratio also presents a rise (although less marked than the Madrid ratio). This increase in the international ratio was detected and termed the "Waldmeier discontinuity" by Svalgaard (2010). In order to homogenize the international series, Svalgaard proposed increasing the *ISN* values prior to 1946 by 20%. However, in evaluating several features of the solar cycle using the standard and the proposed adjusted *ISN*, Aparicio, Vaquero & Gallego (2012) find that the proposed index with the increased values did not significantly improve the characteristics studied.

Returning to the Madrid ratio (Figure 4), this unstable and well-delimited behaviour in a certain time interval (1946-1972) requires explanation. In the documentary sources that we have gathered, there is no mention of any change in instrumentation or methods that can cause such a scaling discontinuity (see Section 2). Neither can it be due to the large number of different observers because the noisy phase is also present in each individual series (Figure 5).

A possible explanation for the unstable phase is that there might have been a change in the quality of the observations. Figure 6 (upper plot) shows the quality of the observations in the fourth period of records. The quality indices were assigned by the original observers at the Madrid Astronomical Observatory. One notes that we detected that three different scales were required to assess observation quality during this period – a scale of 1-3 during the first subperiod (1936-1947), of 1-4 during the second (1948-1980), and of 1-6 during the last (1981-1986). In each case, the best observation quality was represented by 1 on the scale. The means of the possible values for each set of observations are 2 for the first subperiod, 2.5 for the second, and 3.5 for the third. That is, for identical observing conditions, the greater the number of quality levels the higher the mean value. However, although the use of different observation quality scales during this period makes interpretation more difficult, it is noticeable that there are no marked jumps in 1946 and 1973. We next unified the different scales in order to get rid of the jumps at the scale changes. In this way, we obtained a more uniform series (Figure 6, lower plot). To carry out this normalization, we used the following simple procedure: we multiplied every value by the mean of the possible values during the first subperiod and divided it by the mean of the possible values during the subperiod to which it belonged. Hence, the normalization factors were as follows: 2/2 for the first subperiod, 2/2.5 for the second, and 2/3.5 for the third. The result, Figure 6 (lower plot), shows that the quality of the observations during the unstable phase (1946-1972) in no way stands out, but roughly follows the same pattern of values and oscillations as the previous and subsequent years. There was therefore no influence of the observation quality during the unstable phase.

Although we thus did not find any reason for the unstable phase, there is evidence suggesting that it is related to a higher count of individual spots. Firstly, for the entire study period, the form in which the ratio oscillates is similar to that of the *MSN* and the *MGSN* individually (Figure 7). Recall that the *MSN*/*MGSN* ratio is an indicator of the spots/groups ratio. The spots/groups ratio therefore increases near the cycle peaks. Hence, if one assumes a higher count of individual spots during the unstable phase, one will have abnormally high values of the ratio near the cycle peaks (see Figure 7). And secondly, one observes in Figure 8 that, for the fourth period, the oscillations of the *MSN*/*MGSN* and *MSN*/*ISN* ratios have the same shape. This means that the higher the spots/groups ratio (indicated by *MSN*/*MGSN*), the greater is *MSN* relative to *ISN*. This therefore suggests that the abnormal values of the *MSN*/*MGSN* ratio are related to a higher count of individual spots during the unstable phase (because when the spots/groups ratio increased then there occurred an overestimate of *MSN* relative to *ISN*).

5. Sunspot Area

Table 1 lists the distribution over the periods of the study of the Madrid Sunspot Area (*MSA*) records retrieved. Note that we shall be referring to areas corrected for foreshortening, and that the minimum area reported is 2 millionths of a solar hemisphere. As in Section 3, monthly means are calculated from daily data, and yearly means from monthly data (monthly means from daily records, or else monthly records, depending on availability).

Figure 9 shows the temporal evolution of the *MSA* index. Comparing it with Figure 2, one observes that it has maxima and minima in the same years as *MGSN* except for the last cycle, in which the minimum occurs one year earlier and the maximum one year later. With respect to the shape of the cycles, except for the first of the *MSA*, the other cycles are similar to those of the *MGSN*. Moreover, in the last period, the proportion of these three cycles is the same as in the *MGSN* case.

Figure 10 is a plot of both the *MSA/MGSN* ratio and *MSA*. Considering the complete solar cycles, one can see that the ratio varies with the cycle, and that its maxima and minima coincide with those of the *MSA*. Recall that in the previous section it was seen that the *MSN/MGSN* ratio indicated that *s/g* oscillated similarly to the solar cycle. Thus, our data indicate that, at the beginning of the solar cycle (the minimum), the average area and the *s/g* ratio have their minima, and they have their maxima at the peak of the cycle. This can be interpreted as being because at the beginning of the cycle there are small spots with no penumbra, and the groups have few spots (hence the low value of the average area and of the *s/g* ratio), and at the peak of the cycle there are larger spots with an extended penumbra, and the groups have many spots (hence the high value of the average area and of the *s/g* ratio). That this interpretation seems logical gives a first idea of the reliability of the Madrid data.

We shall now compare the *MSA* series with that constructed by Balmaceda et al. (2009), the *BSA* series. To examine the reliability of the *MSA*, we made a linear least-squares fit taking as dependent variable the monthly *MSA* and as independent variable the monthly *BSA* (Figure 11). In evaluating the first period, we found only a moderate correlation ($r = 0.783$), which then increased in the second period ($r = 0.957$) and even more so in the last period ($r = 0.971$). The correlation was also strong for the whole period ($r = 0.963$). The statistical significance corresponded to a confidence level greater than 99.9% for all the aforementioned fits. Thus, only the period 1952-1986 can be considered reliable, because the first period did not show a strong correlation, and the second period is really too short for any appropriate statistical analysis.

The following are the equations of the linear fits corresponding to the different periods (and the total period):

1914-1920:  $MSA = (0.50 \pm 0.04)BSA + (181 \pm 44)$

1931-1933:  $MSA = (1.07 \pm 0.07)BSA + (-9 \pm 13)$

1952-1986:  $MSA = (0.78 \pm 0.01)BSA + (-48 \pm 14)$

1914-1986:  $MSA = (0.76 \pm 0.01)BSA + (-27 \pm 13)$

Lastly, we calculated the scaling factor for each period using the "bisector line" method described by Balmaceda et al. (2009) (Table 3). Thus, each value of the *MSA* series can be compared with its corresponding value in the *BSA* series (or even fill in the gaps in this latter series).

**6. Conclusions**

We have recovered daily Sunspot Number and Group Sunspot Number data for the time spans 1876-1896 and 1936-1986, and monthly Group Sunspot Number data for the time spans 1906-1920, 1931-1933, and 1935, recorded at the Madrid Observatory. We also recovered daily Sunspot Area data for the time spans 1914-1920 and 1952-1986, and monthly data for the time span 1931-1933, recorded at that same Observatory. Our analysis of that data showed that the *MSN*, the *MGSN*, and the last period of the *MSA* can be considered as reliable series because they are strongly correlated with their respective international series.

The spots/groups ratio was evaluated via the *MSN/MGSN* ratio. It presented stable values except during the period of 1946-1972 when the values were higher and there were strong irregular variations. We checked that changes in instrumentation, methods, observers, or quality of the observations do not seem to be the cause of this anomaly. Although these new recovered data from the Madrid Observatory did not allow any firm conclusions to be drawn, they did provide clues that this unstable phase may be related to a higher count of individual spots. It is therefore necessary to seek identical patterns to those found in the Madrid series in other series covering this same time span. Even though there might exist higher counts of individual spots during the unstable phase, there appears to be no concomitant increase in the sunspot area. A study of the size of the individual spots during the unstable phase might help resolve this issue, because there may have occurred an increase in the number of small spots during this time span, i.e., the converse of what was found by Lefèvre & Clette (2011) for solar cycle 23. Such an excess in small spots could be related to what Svalgaard (2010) called the "Waldmeier discontinuity". In particular, that author found a roughly 20% increase starting from 1946 in the *ISN* relative to other indices such as the Diurnal Geomagnetic Variation, the Royal Greenwich Observatory Sunspot Area, Ca-K spectroheliograms, and the Ionospheric Critical Frequency. By definition, an increase in the number of small spots would affect the *ISN* far more strongly (all spots contribute with equal weight) than the other indices which are based on total areas or fluxes, and which are strongly weighted in favour of large spots.

**Acknowledgements**

We wish to acknowledge the assistance provided by the Library and Archive staff of the "Centro Territorial de la AEMET" (Badajoz, Spain), the "Observatorio Astronómico de Madrid" (Madrid, Spain), the "Observatório Astronómico de Lisboa" (Lisbon, Portugal), and the "Real Observatorio de la Armada" (San Fernando, Spain). J.M. Vaquero benefited from the impetus from participating in the Sunspot Number Workshops (http://ssnworkshop.wikia.com/wiki/Home). Support from the Junta de Extremadura (Research Group Grant No. GR10131), from the Ministerio de Economía y Competitividad of the Spanish Government (AYA2011-25945), and from the COST Action ES1005 TOSCA (http://www.tosca-cost.eu) is gratefully acknowledged.

**Appendix**

Bibliographic references of the papers containing original sunspot number and area data from the solar observations of the Madrid Astronomical Observatory are listed in this appendix. We use the following abbreviations: *Anales de la Sociedad Española de Física y Química* (ASEFQ), *Anuario del Observatorio Astronómico de Madrid* (AOAM), *Astronomischen Mitteilungen von Rudolf Wolf* (AMRW) and *Boletín Astronómico del Observatorio de Madrid* (BAOM).

First period

Ventosa, V.: 1876, *Astr. Nach.* **88**, 285-286 and 299-300.

Second period

Third period

Carrasco, R., and Aguilar Stuyck, M.: 1932, *BAOM* **I(2)**, 7-8.
Carrasco, R., and Aguilar Stuyck, M.: 1932, *BAOM* **I(3)**, 6-8.
Carrasco, R., and Aguilar Stuyck, M.: 1932, *BAOM* **I(5)**, 5-6.
Carrasco, R., and Aguilar Stuyck, M.: 1932, *BAOM* **I(7)**, 5-6.
Carrasco, R., and Aguilar Stuyck, M.: 1932, *BAOM* **I(8)**, 8.
Carrasco, R., and Aguilar Stuyck, M.: 1932, *BAOM* **I(9)**, 7-8.
Carrasco, R., and Aguilar Stuyck, M.: 1933, *BAOM* **I(10)**, 7-8.
Carrasco, R., and Santaló, M.: 1933, *BAOM* **I(11)**, 7-8.
Carrasco, R., and Santaló, M.: 1933, *BAOM* **I(14)**, 5-6.
Carrasco, R., and Santaló, M.: 1933, *BAOM* **I(15)**, 5-7.
Carrasco, R.: 1934, *BAOM* **I(18)**, 4-6.
Carrasco, R.: 1934, *BAOM* **I(19)**, 4-6.
Carrasco, R.: 1934, *BAOM* **I(20)**, 4-6.

Fourth period

Gullón, E.: 1936, *Revista de la Academia de Ciencias Exactas, Físicas y Naturales de Madrid* **XXXIII(1)**, 88-91.
Gullón, E.: 1936, *BAOM* **II(4)**, 33-34.
Gullón, E., and Martín Lorón, M.: 1941, *BAOM* **II(7)**, 42-47.
Gullón, E.: 1942, *BAOM* **II(8)**, 45-52.
Gullón, E.: 1943, *BAOM* **II(10)**, 22-28.
Gullón, E., and Martín Lorón, M.: 1944, *BAOM* **III(1)**, 14-16.
Gullón, E.: 1944, *BAOM* **III(2)**, 22-43.
Gullón, E.: 1945, *BAOM* **III(3)**, 1-23.
Gullón, E.: 1946, *BAOM* **III(4)**, 1-26.
Gullón, E.: 1946, *BAOM* **III(6)**, 43-67.
Gullón, E.: 1947, *BAOM* **III(7)**, 34-60.
Gullón, E.: 1948, *BAOM* **III(8)**, 33-61.
Gullón, E.: 1949, *BAOM* **IV(1)**, 33-62.
Gullón, E.: 1950, *BAOM* **IV(2-3)**, 45-71 and 115-142.
Gullón, E.: 1951, *BAOM* **IV(4)**, 39-67.
Gullón, E.: 1952, *BAOM* **IV(5)**, 45-72.
Gullón, E.: 1953, *BAOM* **IV(6)**, 41-49.
Gullón, E.: 1954, *BAOM* **IV(7)**, 47-106.
Gullón, E.: 1955, *BAOM* **IV(8)**, 45-101.
Gullón, E., and López Arroyo, M.: 1956, *BAOM* **V(1)**, 47-117.
Gullón, E., and López Arroyo, M.: 1957, *BAOM* **V(2)**, 61-216.

Table 1. Recovered data sorted by type and period. It should be noted that in 1935 there are only monthly records of the number of groups.

|  | **First period (1876-1896)** | **Second period (1906-1920)** | **Third period (1931-1933)** | **Fourth period (1935-1986)** |
|---|---|---|---|---|
| **Groups** | Daily | Monthly | Monthly | Daily |
| **Spots** | Daily | - | - | Daily |
| **Area** | - | Daily[*] | Monthly | Daily[+] |

[*]Only for 1914-1920.

[+]Only for 1952-1986.

Table 2. Yearly averages of group count (in parentheses, the number of observed days for each year). Each row corresponds to a decade, and each column to the last digit of each year.

|      | 0 | 1 | 2 | 3 | 4 | 5 | 6 | 7 | 8 | 9 |
|---|---|---|---|---|---|---|---|---|---|---|
| **1870** | - | - | - | - | - | - | 1.1 (267) | 1.0 (308) | 0.4 (288) | 0.7 (277) |
| **1880** | 2.9 (244) | 5.0 (278) | 5.5 (286) | 5.9 (237) | 5.9 (253) | 4.7 (232) | 2.3 (214) | 1.5 (287) | 0.9 (267) | 0.6 (284) |
| **1890** | 1.0 (298) | 3.7 (260) | 6.5 (219) | 7.3 (213) | 6.5 (129) | 5.4 (161) | 3.7 (166) | - | - | - |
| **1900** | - | - | - | - | - | - | 2.3 (303) | 4.6 (195) | 5.1 (211) | 3.7 (249) |
| **1910** | 2.4 (278) | 0.8 (263) | 0.4 (305) | 0.3 (276) | 1.6 (131) | 5.2 (224) | 6.7 (236) | 9.6 (221) | 7.4 (211) | 4.7 (163) |
| **1920** | 2.9 (187) | - | - | - | - | - | - | - | - | - |
| **1930** | - | 2.1 (53) | 1.4 (242) | 0.7 (193) | - | 2.8 (263) | 6.3 (226) | 8.6 (273) | 8.8 (246) | 6.6 (73) |
| **1940** | 4.8 (211) | 3.8 (241) | 2.1 (229) | 1.1 (310) | 0.8 (348) | 2.8 (336) | 6.5 (287) | 9.4 (237) | 8.3 (269) | 8.5 (268) |
| **1950** | 5.1 (277) | 4.0 (259) | 1.9 (265) | 0.9 (281) | 0.3 (311) | 2.5 (242) | 8.6 (281) | 11.2 (98) | 11.4 (165) | 9.5 (116) |
| **1960** | 6.6 (82) | 3.0 (168) | 2.0 (210) | 1.5 (189) | 0.5 (219) | 0.8 (252) | 2.5 (276) | 5.3 (332) | 5.7 (316) | 5.2 (287) |
| **1970** | 5.3 (301) | 4.3 (272) | 3.8 (229) | 2.2 (264) | 2.3 (260) | 0.9 (278) | 0.8 (252) | 1.9 (217) | 6.0 (232) | 8.6 (195) |
| **1980** | 8.1 (216) | 8.1 (206) | 6.4 (182) | 4.2 (175) | 2.7 (169) | 1.0 (170) | 0.7 (120) | - | - | - |

Table 3. Scaling factor, correlation coefficient, and statistical significance for each period of the *MSA* series.

|                 | **1914-1920** | **1931-1933** | **1952-1986** |
|-----------------|---------------|---------------|---------------|
| **Scal. Factor**    | 1.516         | 0.942         | 1.317         |
| **Corr. Coeff.**    | 0.615         | 0.923         | 0.958         |
| **Stat. Sign. (%)** | 99.9          | 99.9          | 99.9          |

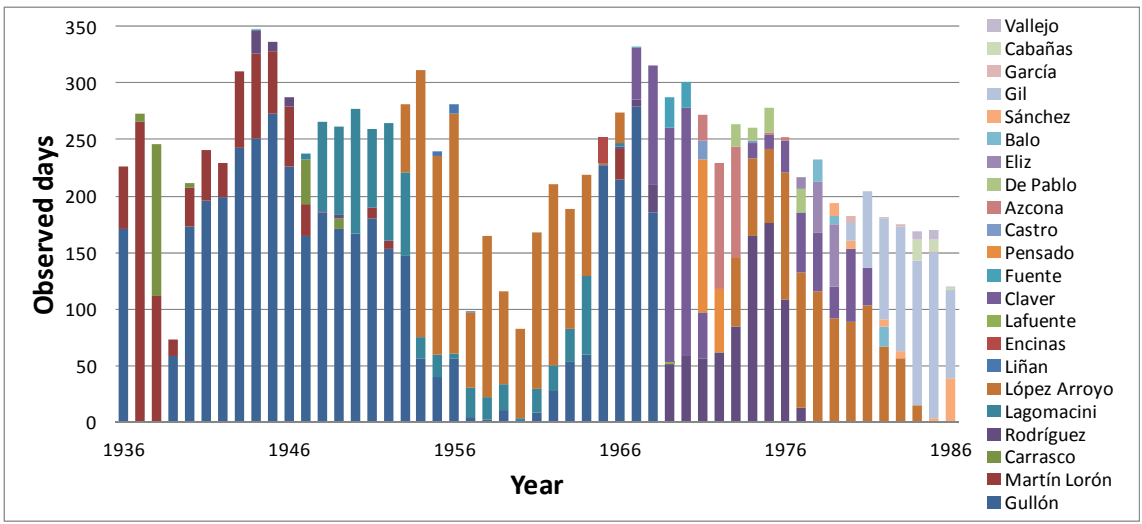

Fig. 1. Days observed per year by each observer from 1936 to 1986.

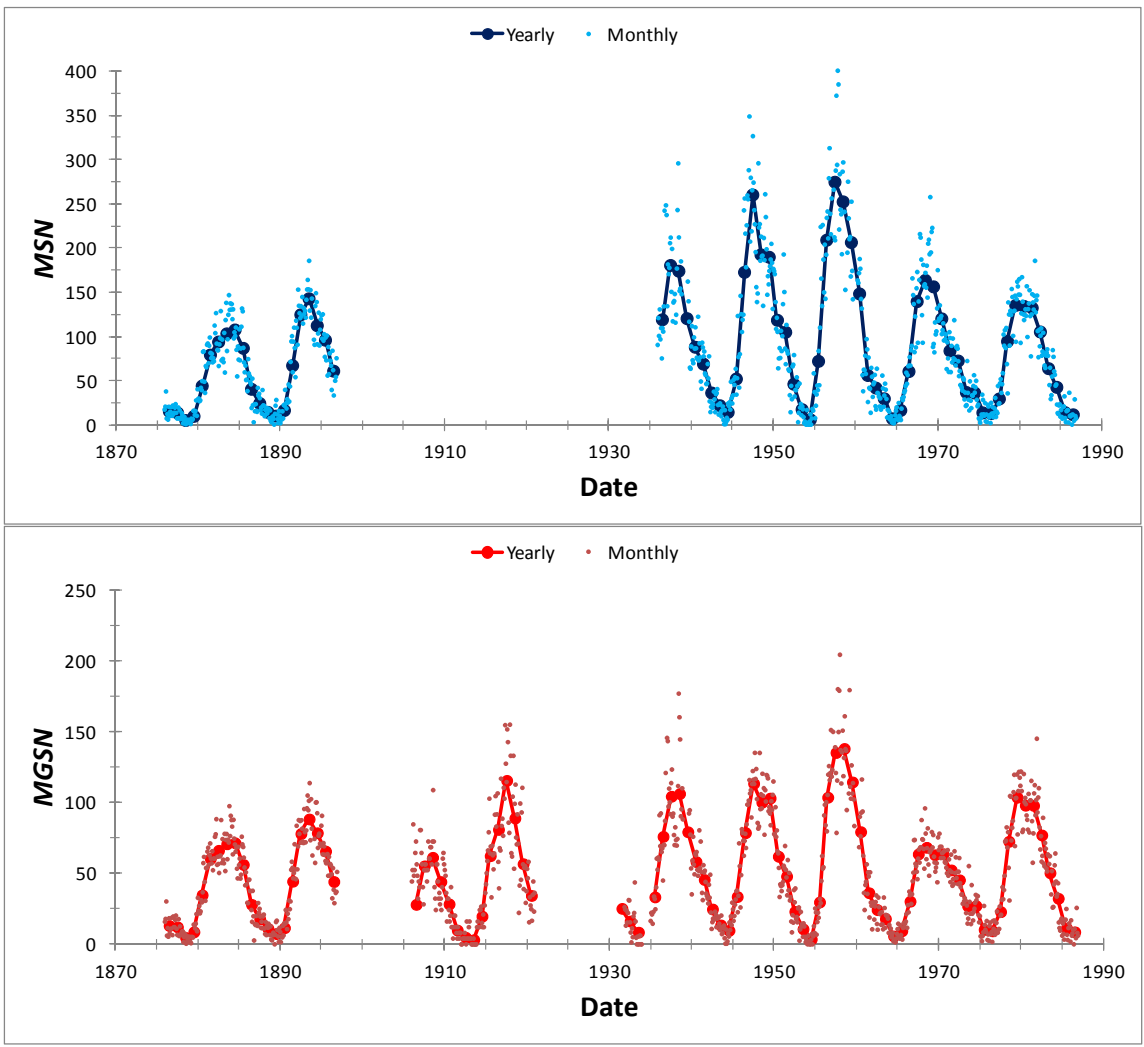

Fig. 2. Monthly and yearly *MSN* (top panel) and *MGSN* (bottom panel) values from 1876 to 1986.

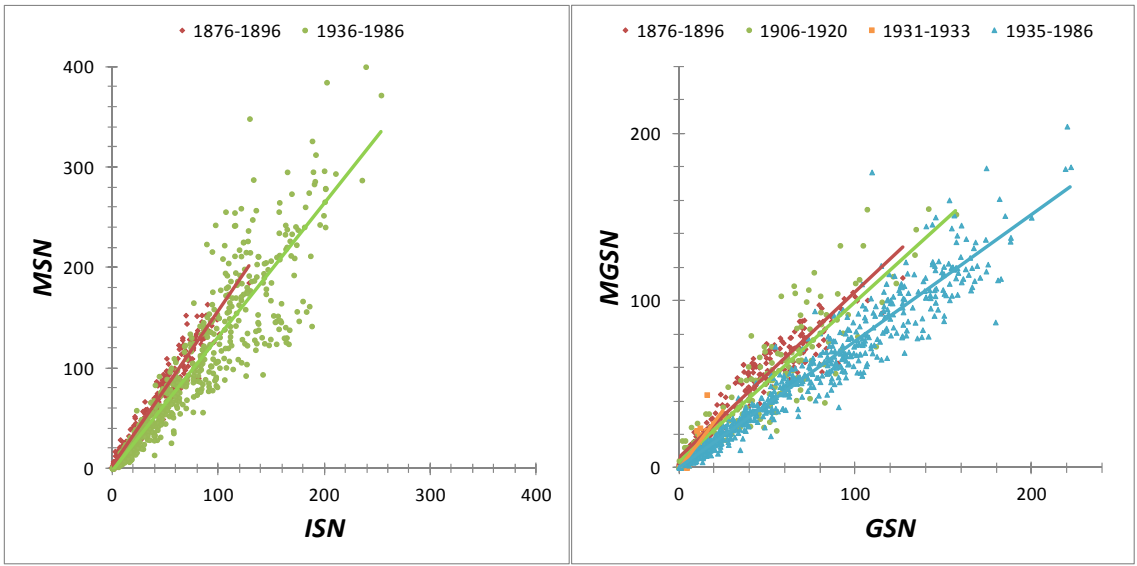

Fig. 3. Scatter plots of the monthly *MSN* values versus the *ISN* ones (left panel) and the monthly *MGSN* values versus the *GSN* ones (right panel).

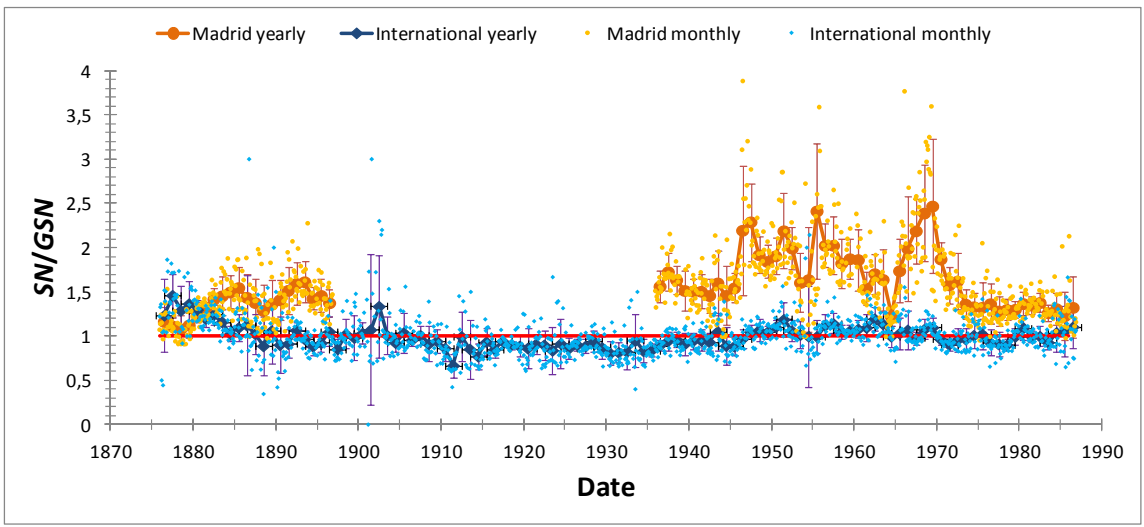

Fig. 4. Monthly and yearly values of the ratios *MSN/MGSN* and *ISN*/*GSN*. Error bars represent one standard deviation.

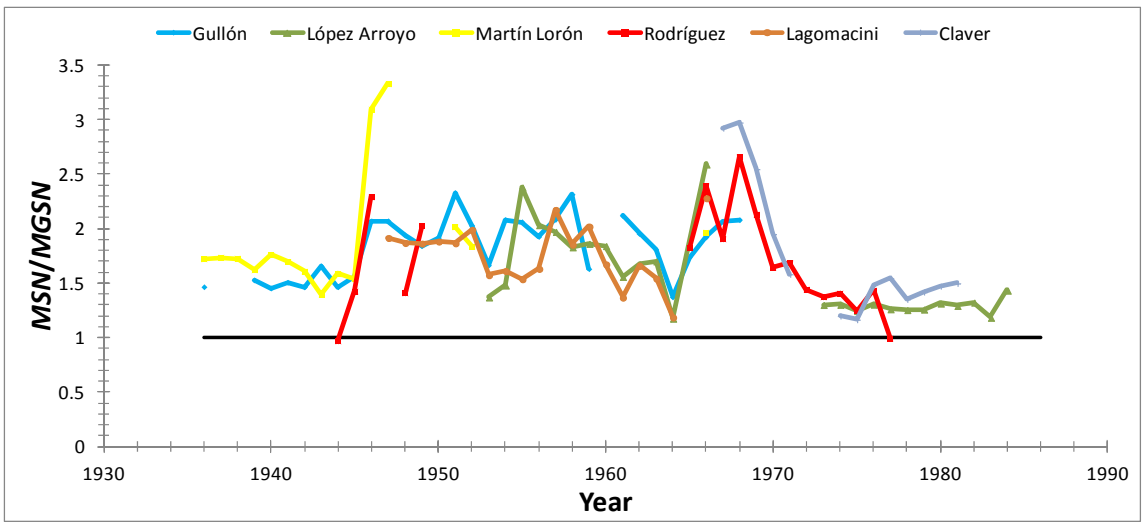

Fig. 5. Yearly values of *MSN/MGSN* from 1936 to 1986 obtained by the main observers.

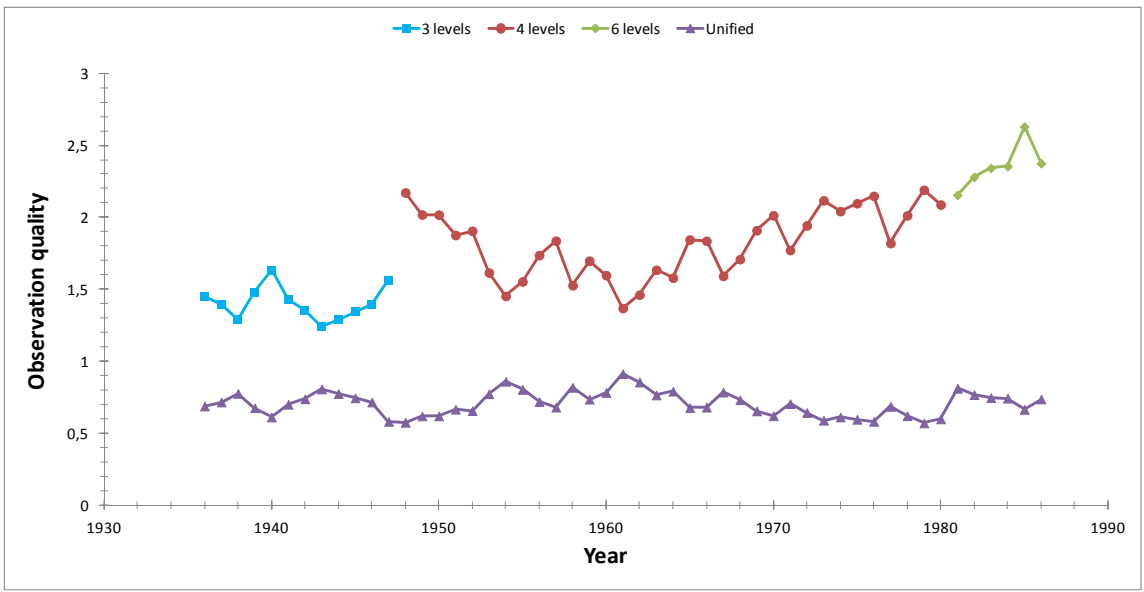

Fig. 6. Yearly average of the observation quality level. The first subperiod (1936-1947) has three quality levels, the second (1948-1980) four levels, and the third (1981-1986) six levels. In each subperiod, the value 1 indicates the highest quality. The purple line represents the yearly average of the observation quality level, normalizing the three subperiods to the same scale. In this case, the inverse of the original values was calculated. Thus, the highest quality of observation corresponds to the highest values.

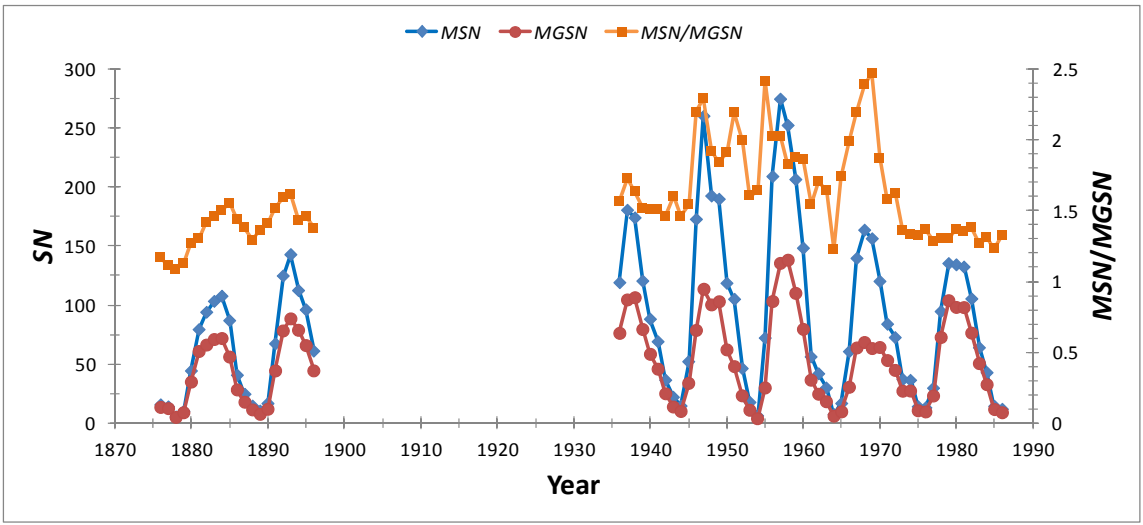

Fig. 7. Yearly values of the *MSN*, the *MGSN*, and the *MSN/MGSN* ratio from 1876 to 1986.

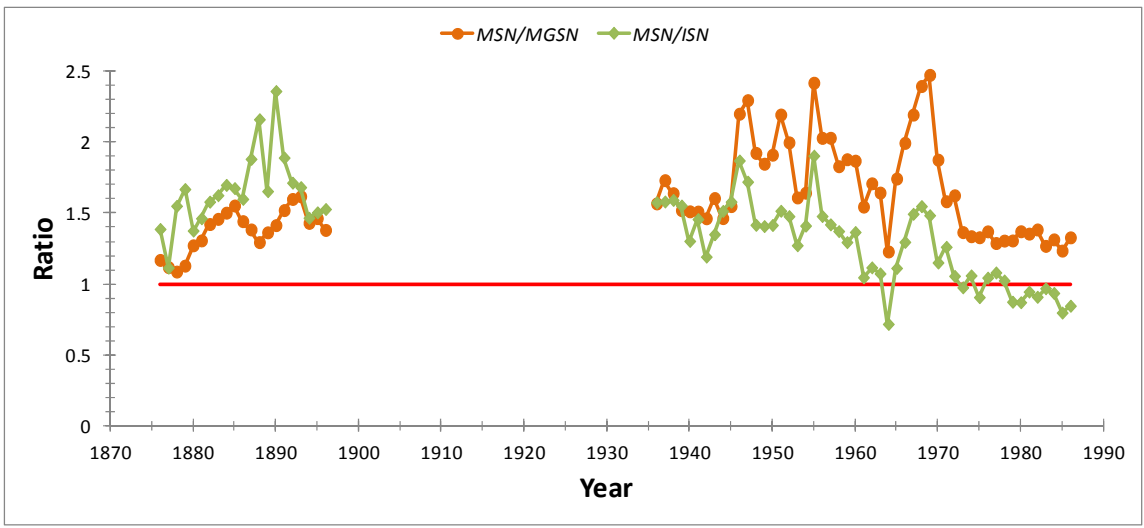

Fig. 8. Yearly values of *MSN*/*MGSN* and *MSN*/*ISN* from 1876 to 1986.

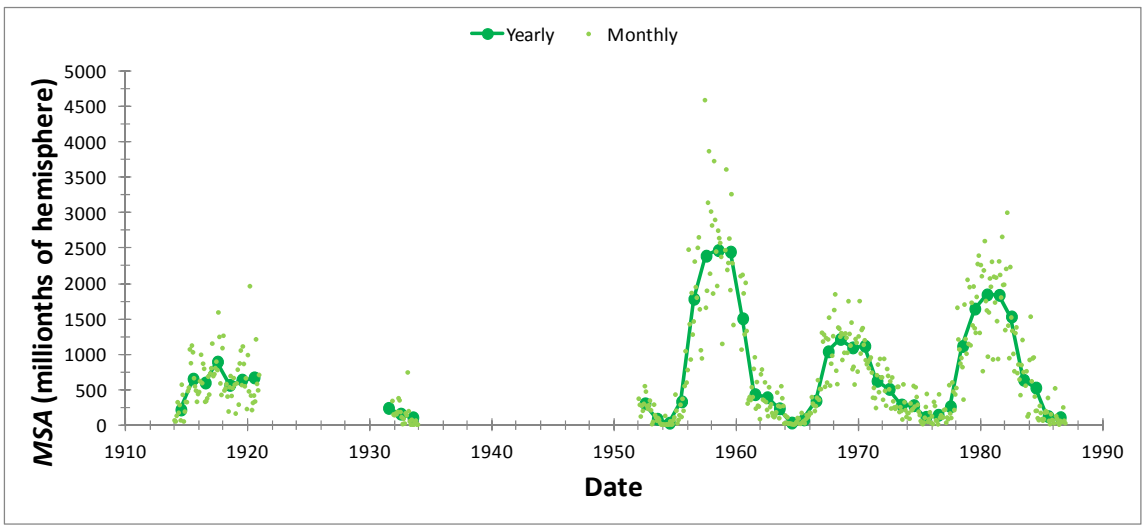

Fig. 9. Monthly and yearly values of the *MSA* from 1914 to 1986.

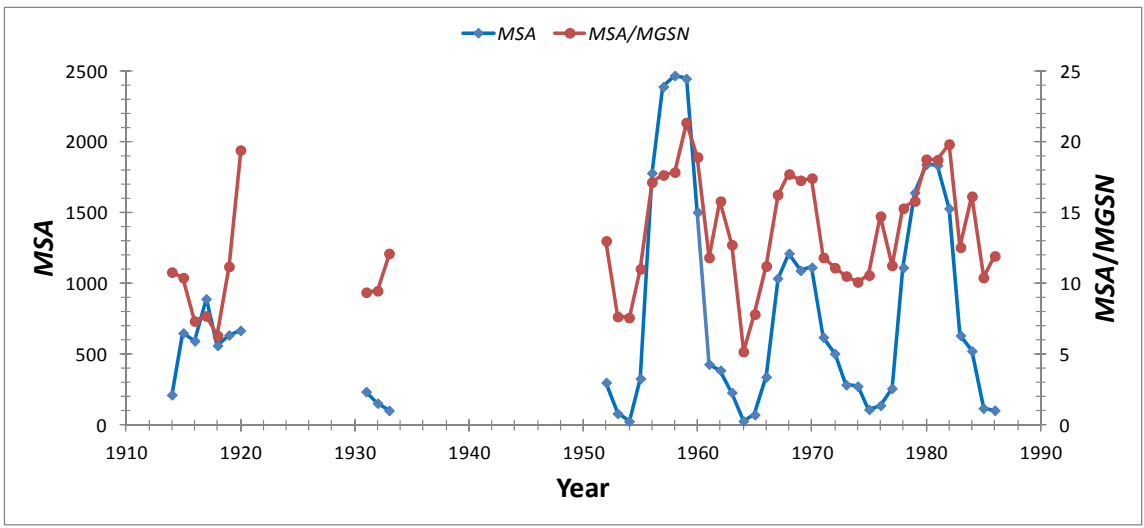

Fig. 10. Yearly values of the *MSA* and the *MSA*/*MGSN* ratio from 1914 to 1986.

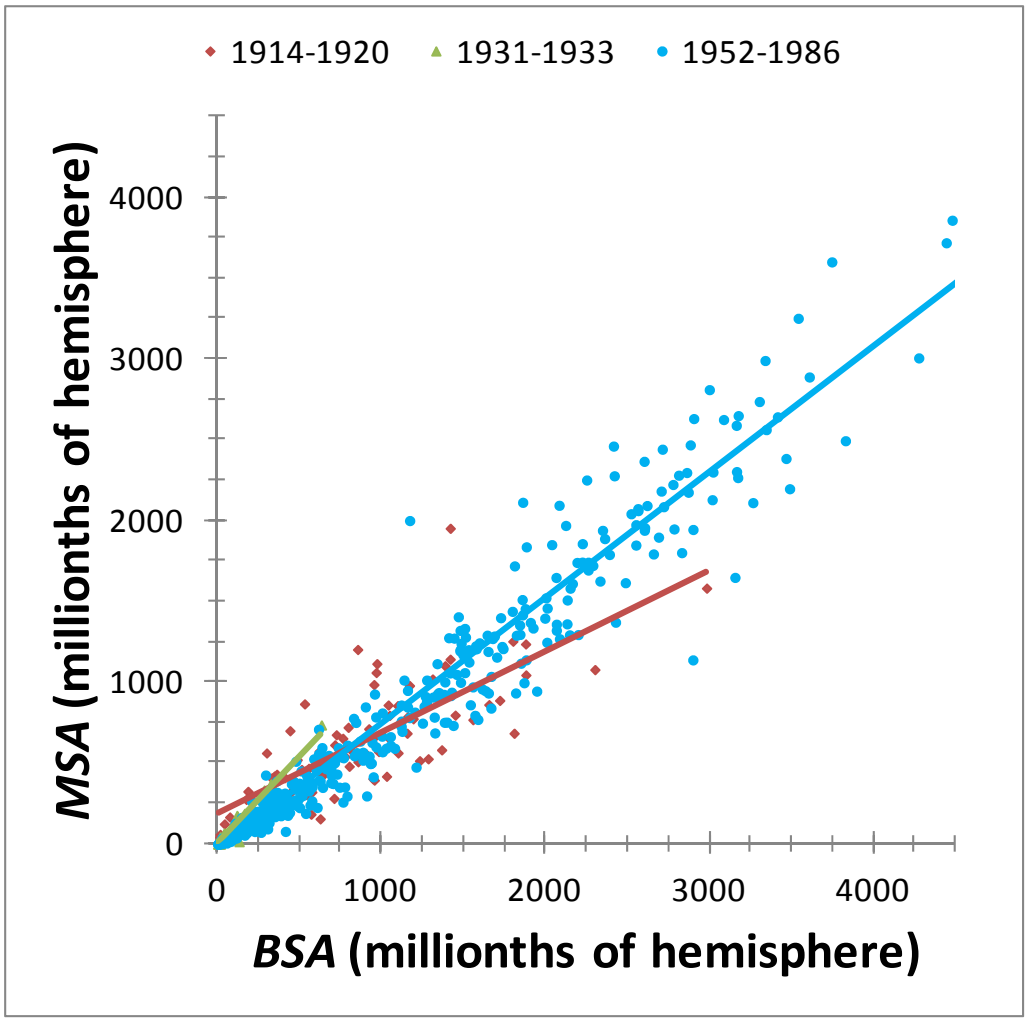

Fig. 11. Scatter plot of the monthly *MSA* values versus the *BSA* ones.